\documentclass[useAMS,usenatbib,usegraphicx]{mn2e}
\usepackage{amssymb}
\usepackage{xcolor}
\usepackage{bm}

\begin{document}

\newcommand{\fixme}[1]{{\textbf{Fixme: #1}}}
\newcommand{\detD}{{\det\!\cld}}
\newcommand{\clh}{\mathcal{H}}
\newcommand{\ud}{{\rm d}}
\newcommand{\eprint}[1]{\href{http://arxiv.org/abs/#1}{#1}}
\newcommand{\adsurl}[1]{\href{#1}{ADS}}
\newcommand{\ISBN}[1]{\href{http://cosmologist.info/ISBN/#1}{ISBN: #1}}
\newcommand{\jcap}{J.\ Cosmol.\ Astropart.\ Phys.}
\newcommand{\mnras}{Mon.\ Not.\ R.\ Astron.\ Soc.}
\newcommand{\physrep}{Phys.\ Rep.}
\newcommand{\progress}{Rep.\ Prog.\ Phys.}
\newcommand{\prlett}{Phys.\ Rev.\ Lett.}
\newcommand{\prd}{Phys.\ Rev.\ D}
\newcommand{\apj}{ApJ}

\newcommand{\aap}{A\&A}
\newcommand{\vort}{\varpi}
\newcommand\ba{\begin{eqnarray}}
\newcommand\ea{\end{eqnarray}}
\newcommand\be{\begin{equation}}
\newcommand\ee{\end{equation}}
\newcommand\lagrange{{\cal L}}
\newcommand\cll{{\cal L}}
\newcommand\cln{{\cal N}}
\newcommand\clx{{\cal X}}
\newcommand\clz{{\cal Z}}
\newcommand\clv{{\cal V}}
\newcommand\cld{{\cal D}}
\newcommand\clt{{\cal T}}

\newcommand{\ThreeJSymbol}[6]{\begin{pmatrix}
#1 & #3 & #5 \\
#2 & #4 & #6
 \end{pmatrix}}

\newcommand\clo{{\cal O}}
\newcommand{\cla}{{\cal A}}
\newcommand{\clp}{{\cal P}}
\newcommand{\clr}{{\cal R}}
\newcommand{\uD}{{\mathrm{D}}}
\newcommand{\calE}{{\cal E}}
\newcommand{\calB}{{\cal B}}
\newcommand{\curl}{\,\mbox{curl}\,}
\newcommand\del{\nabla}
\newcommand\Tr{{\rm Tr}}
\newcommand\half{{\frac{1}{2}}}
\newcommand\fourth{{1\over 8}}
\newcommand\bibi{\bibitem}
\newcommand{\kf}{\beta}
\newcommand{\kfprod}{\alpha}
\newcommand\calS{{\cal S}}
\renewcommand\H{{\cal H}}
\newcommand\K{{\rm K}}
\newcommand\mK{{\rm mK}}
\newcommand\km{{\rm km}}
\newcommand\synch{\text{syn}}
\newcommand\opacity{\tau_c^{-1}}

\newcommand{\Psil}{\Psi_l}
\newcommand{\bI}{\bar{I}}
\newcommand{\bq}{\bar{q}}
\newcommand{\bv}{\bar{v}}
\renewcommand\P{{\cal P}}
\newcommand{\numfrac}[2]{{\textstyle \frac{#1}{#2}}}

\newcommand{\Omtot}{\Omega_{\mathrm{tot}}}
\newcommand\xx{\mbox{\boldmath $x$}}
\newcommand{\phpr} {\phi`}
\newcommand{\gam}{\gamma_{ij}}
\newcommand{\sqgam}{\sqrt{\gamma}}
\newcommand{\delk}{\Delta+3{\K}}
\newcommand{\dph}{\delta\phi}
\newcommand{\om} {\Omega}
\newcommand{\dom}{\delta^{(3)}\left(\Omega\right)}
\newcommand{\rar}{\rightarrow}
\newcommand{\Rar}{\Rightarrow}
\newcommand\gsim{ \lower .75ex \hbox{$\sim$} \llap{\raise .27ex \hbox{$>$}} }
\newcommand\lsim{ \lower .75ex \hbox{$\sim$} \llap{\raise .27ex \hbox{$<$}} }
\newcommand\bigdot[1] {\stackrel{\mbox{{\huge .}}}{#1}}
\newcommand\bigddot[1] {\stackrel{\mbox{{\huge ..}}}{#1}}
\newcommand{\Mpc}{\text{Mpc}}
\newcommand{\Al}{{A_l}}
\newcommand{\Bl}{{B_l}}
\newcommand{\eAl}{e^\Al}
\newcommand{\ix}{{(i)}}
\newcommand{\ixp}{{(i+1)}}
\renewcommand{\k}{\beta}
% Derivatives
\newcommand{\HD}{\mathrm{D}}

\newcommand{\nonflat}[1]{#1}
\newcommand{\Cgl}{C_{\text{gl}}}
\newcommand{\Cgltwo}{C_{\text{gl},2}}
\newcommand{\He}{{\rm{He}}}
\newcommand{\Mhz}{{\rm MHz}}
\newcommand{\vx}{{\mathbf{x}}}
\newcommand{\ve}{{\mathbf{e}}}
\newcommand{\vv}{{\mathbf{v}}}
\newcommand{\vk}{{\mathbf{k}}}
\newcommand{\vn}{{\mathbf{n}}}

\newcommand{\vnhat}{{\hat{\mathbf{n}}}}
\newcommand{\vkhat}{{\hat{\mathbf{k}}}}
\newcommand{\taueps}{{\tau_\epsilon}}

\newcommand{\vgrad}{{\mathbf{\nabla}}}
\newcommand{\fbarln}{\bar{f}_{,\ln\epsilon}(\epsilon)}

%%%%%%%%%%%%%%%%%%%%%%%%%%%%%%%%%%%%%%%%%%%%%%%%

\title[Intrinsic alignments in lensing cross-correlations]{Intrinsic alignments in the cross-correlation of cosmic shear and CMB weak lensing}

\author[Alex Hall and Andy Taylor]
{Alex Hall\thanks{ahall@roe.ac.uk} and Andy Taylor\\
Institute for Astronomy, University of Edinburgh, Royal Observatory, Blackford Hill, Edinburgh, EH9 3HJ, U.K.}
\maketitle
%\email{ahall@roe.ac.uk}
%\affiliation{Institute for Astronomy, University of Edinburgh, Royal Observatory, Blackford Hill, Edinburgh, EH9 3HJ, U.K.}

%\author{Andy Taylor}
%\email{ant@roe.ac.uk}
%\affiliation{Institute for Astronomy, University of Edinburgh, Royal Observatory, Blackford Hill, Edinburgh, EH9 3HJ, U.K.}

\begin{abstract}
We demonstrate that the intrinsic alignment of galaxies with large-scale tidal fields sources an extra contribution to the recently-detected cross-correlation of galaxy shear and weak lensing of the microwave background. The extra term is the analogy of the `GI' term in standard cosmic shear studies, and results in a reduction in the amplitude of the cross-correlation. We compute the intrinsic alignment contribution in linear and non-linear theory, and show that it can be at roughly the 15\% level for the CFHT Stripe 82 redshift distribution, if the canonical amplitude of intrinsic alignments is assumed. The new term can therefore potentially reconcile the apparently low value of the measured cross-correlation with standard $\Lambda$CDM. We discuss various small-scale effects in the signal and the dependence on the source-redshift distribution. We discuss the exciting possibility of self-calibrating intrinsic alignments with a joint analysis of cosmic shear and weak lensing of the microwave background.
\end{abstract}

\maketitle

\section{Introduction}
\label{sec:intro}

In recent years, gravitational lensing of radiation by large-scale structure in the Universe has proved a fruitful probe of cosmology. By measuring the characteristic distortion of sources across the sky and as a function of redshift, estimators can be constructed that constrain cosmological parameters~\citep{2013arXiv1303.5077P, 2013MNRAS.432.2433H}. The advantage of this technique is that it is sensitive to the total matter distribution, and is thus insensitive to the uncertain clustering bias of visible tracers.

The two most promising avenues for using this technique are weak lensing of the cosmic microwave background (CMB; see, e.g.~\citealt{2006PhR...429....1L} for a review), and weak lensing of galaxies, (e.g.~\citealt{2001PhR...340..291B}). The first approach makes use of the statistical anisotropy imparted on the CMB temperature and polarization fluctuations for a given realisation of the lensing field. Quadratic estimators for CMB weak lensing can then be constructed from the observed CMB maps~\citep{2003PhRvD..67h3002O}. In contrast, weak lensing of individual galaxies is probed by measuring the statistics of the shear field, which imparts coherent distortions into the shapes of galaxies. Given knowledge of the unlensed `intrinsic' shape distribution, the statistics of galaxy ellipticities may be used to constrain cosmology. Two-point statistics for both these effects have been measured, the state of the art being the \emph{Planck} satellite for CMB weak lensing~\citep{2013arXiv1303.5077P} and the Canada-France-Hawaii Legacy Survey (CFHTLenS;~\citealt{2013MNRAS.432.2433H}) and Deep Lens Survey (DLS;~\citealt{2013ApJ...765...74J}) for cosmic shear.

To maximise the information extracted from CMB lensing and cosmic shear, we must also consider their cross-correlation. This is expected to be non-zero since any material lensing a background source-galaxy population will also lens CMB photons sourced at much higher redshift~\citep{2002PhRvD..65b3003H}. The cross-power spectrum was recently detected for the first time by combining CMB maps from the Atacama Cosmology Telescope and galaxy lensing data from the CFHT Stripe 82 Survey (CS82;~\citealt{2013arXiv1311.6200H}), with a detection significance of 3.2$\sigma$. The best-fit amplitude for the cross-correlation, relative to the fiducial cosmological model favoured by \emph{Planck} was found to be 0.61 $\pm$ 0.19 with $1\sigma$ errors, i.e. 2$\sigma$ lower than might be expected. The statistical significance of this discrepancy is however marginal, and could be due to uncertainties in the source-galaxy distribution (see~\citealt{2013arXiv1311.6200H} for further discussion).

In this letter, we identify an effect not accounted for by the analysis of~\citet{2013arXiv1311.6200H} which could explain at least some of the discrepancy with the \emph{Planck} and WMAP9 models, that of intrinsic alignments (IAs). The observed ellipticity of a galaxy has contributions from both the lensing shear and the intrinsic ellipticity. Physically close pairs of galaxies are expected to be aligned to some extent, since they both presumably formed in the same tidal field sourced by the surrounding dark matter. This leads to a contribution from IAs in the auto-correlation of observed ellipticities. IAs are a significant source of systematic confusion for lensing studies, and much effort has been expended on their mitigation~\citep[see, e.g.][and references therein]{2012MNRAS.424.1647K}.

If the intrinsic ellipticity of a galaxy is sourced by the gravitational field of its surroundings, a further contribution to ellipticity power spectra arises from the correlation of this field with the shear of a background galaxy at higher redshift, as first pointed out in~\citet{2004PhRvD..70f3526H}. This paper also noted that the same effect would be present in the cross-correlation of CMB lensing and galaxy shear. Furthermore, the relative importance of this term in standard galaxy shear studies can be large if the background source redshift is large compared to the redshift of the intrinsically aligned galaxy~\citep{2010A&A...523A...1J}, as is the case in the correlation with CMB lensing.

We emphasise that given the marginal nature of the discrepancy reported in~\citet{2013arXiv1311.6200H}, the aim of this work is not purely to `solve' a problem that does not exist. Rather, we wish to elucidate and quantify the effect of IAs, demonstrating that they must be accounted for when extracting cosmological constraints from the CMB-shear lensing signal.

%In Section ~\ref{sec:IAs} we compute the power spectrum of the IA contribution to the cross-correlation measured in~\citet{2013arXiv1311.6200H}, using the simple models usually constrained in studies of IAs. We show that it has the correct sign to mitigate the discrepancy with the fiducial model. We recalculate the cross-correlation including this term, and show that the magnitude of the new term is roughly 15\% of the `standard' term, if the canonical value for the amplitude of  IAs is assumed. We consider the importance of non-linear effects on IAs, and briefly discuss the redshift-dependence of the new term. We take $c=1$ throughout.

In Section~\ref{sec:IAs} we introduce the simple IA model used in this work, and in Section~\ref{sec:cls} we compute the power spectrum of the IA contribution to the cross-correlation measured in~\citet{2013arXiv1311.6200H}. We present our results in Section~\ref{sec:results} and conclude in Section~\ref{sec:concs}. We take $c=1$ throughout.

\section{Intrinsic Alignments}
\label{sec:IAs}

We will model IAs with the popular linear alignment model of~\citet{2004PhRvD..70f3526H}, first introduced in~\citet{2001MNRAS.320L...7C}. In this model, the intrinsic shear of a galaxy is assumed to be directly proportional to the large-scale tidal field at the time of galaxy formation. The canonical formulation of this relationship as introduced in~\citet{2004PhRvD..70f3526H} holds that the intrinsic complex shear in a line-of-sight direction $\vnhat$ and comoving distance $r$ is
\begin{equation}
\gamma^{I}(\vnhat,r) = -A\frac{C_1}{4\pi G}\frac{\eth \eth \phi_p(r \vnhat)}{r^2},
\label{eq:lin-IAs}
\end{equation}
where $\eth$ is a spin-raising operator, $\phi_p$ is the Newtonian-gauge gravitational potential at the time of galaxy formation (taken to be well before dark energy domination), $C_1 = 5\times 10^{-14} h^{-2} M_{\odot}^{-1} \mathrm{Mpc}^{3}$, and $A$ is a dimensionless constant. Note that throughout this work we assume that the two Newtonian gauge potentials are equal. A value of $A=1$ was found in~\citet{2007NJPh....9..444B} by matching to the SuperCOSMOS survey~\citep{2002MNRAS.333..501B}. As shown in~\citet{2013MNRAS.432.2433H}, the amplitude parameter $A$ depends strongly on galaxy type, with large values ($4 \lesssim A \lesssim 6$) measured for early-type galaxies and small values, consistent with zero, found for late-type galaxies. Throughout this work, we will assume the canonical value $A=1$, but the appropriate scaling of our results with this parameter should be borne in mind.
%
%The negative sign in Eq.~\eqref{eq:lin-IAs} ensures that the galaxy is aligned with the `stretching' axis of the tidal field (see~\citet{2004PhRvD..70f3526H} for further details).

In the original model of~\citet{2004PhRvD..70f3526H}, $\phi_p$ was taken as the linear potential, and filtered to null the contribution to IAs from scales $k > 1 h \mathrm{Mpc}^{-1}$, where the model of equation~(\ref{eq:lin-IAs}) was not expected to hold. In~\citet{2007NJPh....9..444B} this filter was not included, and $\phi_p$ was replaced by its non-linear value. Note that the use of the non-linear $\phi_p$ lacks strong physical motivation, and its main utility lies in the improved fit to existing data it can provide. On large scales, the model is equivalent to the linear IA model, which has a more convincing scale-dependence although still lacking in a compelling prediction for the redshift-dependence of $A$. The cut-off on small scales has more physical motivation, since it is expected that the IA mechanism will transition from large-scale tidal alignments to inter-halo tidal alignments and torquing \citep{2010ApJ...721..939P, 2012MNRAS.421.2751S}. In this work, we will consider both linear and non-linear IAs, with and without a cut-off on small scales. Note that in choosing this set of models we do not advocate their employment in extracting precise cosmological constraints, since the accuracy of the models is expected to be poor in the non-linear regime where most of the information from lensing is contained. Rather, our choice allows us to demonstrate clearly the utility of IAs in \emph{testing} these models. For example, a more refined IA model would account for the time-dependence of the amplitude $A$, and these kinds of refinements should be considered when forecasting parameter constraints.

Expanding $\phi_p$ in spherical harmonics, we have
\begin{equation}
\gamma^{I}(\vnhat,r) = -A\frac{C_1}{4\pi G} \sum_{lm} \frac{\phi_{p,lm}(r)}{r^2} \sqrt{\frac{(l+2)!}{(l-2)!}}{}_2Y_{lm}(\vnhat),
\label{eq:E-mode}
\end{equation}
where ${}_2Y_{lm}$ is a spin-2 spherical harmonic (see, e.g.~\citealt{1997PhRvL..78.2054S}). Since the $\phi_{lm}$ coefficients have electric-type parity, this is a pure E-mode expansion, as expected for scalar modes. The E-mode coefficients can now be read-off from equation~(\ref{eq:E-mode}). Note that a B-mode component might also be expected from a more complicated IA model.

\section{Angular power spectra}
\label{sec:cls} 

For a single source at comoving distance $r_s$, the cosmic shear field is given by
$\gamma(\vnhat;r_s) = \frac{1}{2}\eth \eth \psi(\vnhat;r_s)$ where the lensing potential $\psi$ is
\begin{equation}
\psi(\vnhat;r_s) = 2\int_0^{r_s} \ud r \; \frac{r_s -r}{r_s r} \phi(\vnhat r, r),
\label{eq:lenspot}
\end{equation}
where we have made the Born approximation and assumed a spatially flat Universe.

The E-mode coefficients of $\gamma$ may now be computed in the same way as for the IA field, and integrated over the distribution of source distances $n_r(r)$. For CMB weak lensing, we assume that the source plane is of zero thickness and at a distance $r_*$, i.e. we assume instantaneous recombination.

It is straightforward to compute the angular power spectra of these fields, and for computational convenience we make the Limber approximation for all our spectra. This should be an excellent approximation on the angular scales we consider for both the lensing and IA kernels, since the distance distribution $n_r(r)$ for the CS82 survey used in~\citet{2013arXiv1311.6200H} is broad compared to the typical spatial scales that contribute to our results. With this assumption, the E-mode angular power spectra of IAs ($I$), CMB lensing ($\gamma_{CMB}$), and cosmic shear ($\gamma_{gal}$) are
\begin{equation}
C_l^{\gamma_{CMB} \gamma_{gal}} = l^4 \int_0^{r_{max}} \ud r \; \frac{g(r)}{r^2}\frac{(1-r/r_*)}{r} P_{\phi \phi}(l/r,r),
\label{eq:cmb-gal-cl}
\end{equation}
\begin{equation}
C_l^{I \gamma_{gal}} = l^2 \int_0^{r_{max}} \ud r \; \frac{g(r)n_r(r)}{r^2} P_{\phi I}(l/r,r),
\label{eq:cmb-I-cl}
\end{equation}
where $r_{max}$ is the background conformal distance corresponding to the maximum redshift in the galaxy survey. $P_{\phi \phi}$ is the three-dimensional power spectrum of the potential, and $P_{\phi I}$ is given by
\begin{equation}
P_{\phi I}(k,r) = -\frac{A C_1}{4\pi G \bar{D}(a)}k^2 P_{\phi \phi}(k,r),
\end{equation}
where $a$ is the scale factor corresponding to distance $r$ in the background, and $\bar{D}$ is a growth factor for the potential, normalised to unity at high redshift.

The lensing kernel is given by
\begin{equation}
g(r) = \int_r^{r_{max}}\ud r' \; n_r(r')\frac{r'-r}{r'r}.
\end{equation}
Note that these equations include the correct factors of $a$ omitted in~\citet{2004PhRvD..70f3526H} and many subsequent works (see also Appendix B of~\citealt{2011A&A...527A..26J}).

In the non-linear modification advocated in~\citet{2007NJPh....9..444B}, the comoving density power spectrum is scaled using HALOFIT~\citep{2003MNRAS.341.1311S}, and then related to $P_{\phi \phi}$ using the Poisson equation. In this work, our non-linear corrections are scaled using the update to HALOFIT detailed in~\citet{2012ApJ...761..152T}, which is accurate at the 5\% level for $k \leq 1 h \mathrm{Mpc}^{-1}$ and at 10\% for $1\leq k \leq 10 h \mathrm{Mpc}^{-1}$.

We use the CS82 redshift distribution used in~\citet{2013arXiv1311.6200H}, given by
\begin{equation}
n_z(z) = N\frac{z^a + z^{ab}}{z^b + c},
\label{eq:zdist}
\end{equation}
with $a=0.531$, $b=7.810$, $c=0.517$, and $N$ chosen to normalise the distribution. We take $z_{max} = 6$, which is sufficiently large such that the computation of $N$ converges. Note that this distribution is itself uncertain, with realistic variations propagating into $\mathcal{O}(10\%)$ variations in the cross-power spectrum, as discussed in~\citet{2013arXiv1311.6200H}.

% We plot this distribution in Fig.~\ref{fig:zdist}.  Note that this distribution is itself uncertain, with realistic variations propagating into $\mathcal{O}(10\%)$ variations in the cross-power spectrum, as discussed in~\citet{2013arXiv1311.6200H}.

%\begin{figure}
%\centering
%\includegraphics[width=\columnwidth]{cfht_zdist.eps}
%\caption{The CS82 redshift distribution used in this work, normalised to unit area.}
%\label{fig:zdist}
%\end{figure}

\section{Results}
\label{sec:results}

%To compute the angular power spectra presented in Section~\ref{sec:cls}, we assume a flat $\Lambda$CDM model with parameters similar to those used in the fiducial models in~\citet{2013arXiv1311.6200H}. Specifically, we take $\Omega_bh^2 = 0.0223$, $\Omega_ch^2 = 0.112$, $h=0.7$, $A_s = 2.1 \times 10^{-9}$, $n_s = 0.96$, and $w=-1$. We also have $\sigma_8 = 0.79$ as a derived parameter. The primordial power spectrum has a pivot scale of $k_0 = 0.05 \mathrm{Mpc}^{-1}$ and we assume three massless neutrinos. We use the Boltzmann code CAMB~\citep{2000ApJ...538..473L} to compute a reference linear power spectrum today which is then evolved in time using linear theory.

To compute the angular power spectra presented in Section~\ref{sec:cls}, we assume a flat $\Lambda$CDM model with parameters given by the best fitting \emph{Planck} + lensing + WP + highL model from~\citet{2013arXiv1303.5076P}. The primordial power spectrum has a pivot scale of $k_0 = 0.05 \mathrm{Mpc}^{-1}$ and we assume three massless neutrinos. We use CAMB~\citep{2000ApJ...538..473L} to compute a reference linear power spectrum today which is then evolved in time using linear theory.

\begin{figure}
\centering
\includegraphics[width=\columnwidth]{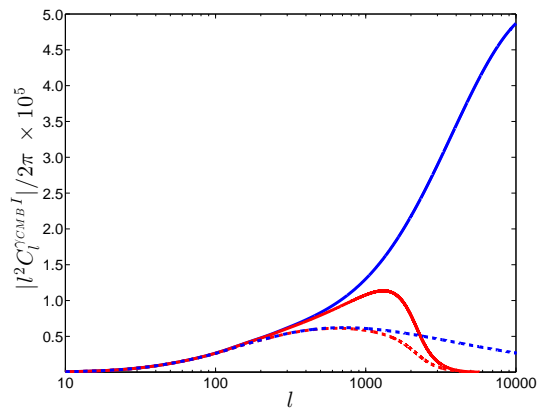}
\caption{The absolute value of the angular cross-power spectrum of the CMB lensing E-mode and IA E-mode, for the source-redshift distribution in equation~(\ref{eq:zdist}) and in the Limber limit. Plotted are the non-linear HALOFIT model (blue, upper solid), the non-linear model with a cut-off on small scales (red, lower solid), the linear theory model (blue, upper dashed), and the linear model with a cut-off (red, lower dashed).}
\label{fig:ia_dcl}
\end{figure}

In Fig.~\ref{fig:ia_dcl} we plot the absolute value of the angular cross-power spectrum of the IA shear E-mode and the CMB lensing shear E-mode, distinguishing between the various small-scale behaviour discussed in Section~\ref{sec:IAs}. The characteristic boosting of the non-linear model over the linear model on small angular scales is clearly seen. The scale dependence of these curves reflects the interplay between the shape of the power spectrum $P_{\phi \phi}$ and the distance kernel in equation~(\ref{eq:cmb-I-cl}). In particular, the turn-over in the linear-theory curve at $l\approx 700$ is due to the turn-over in the matter power spectrum. We have truncated this plot at $l=10000$, since at higher $l$ the integral picks up significant contributions from wavenumbers $k \gtrsim 10 h \mathrm{Mpc}^{-1}$, where our HALOFIT correction is not expected to be accurate. It should also be borne in mind that our IA models are not expected to be accurate on these scales.

Note that the cross-power spectrum is negative, i.e. an \emph{anti}-correlation. The physical reason for this is as follows. Consider as a toy model a density field perturbed at linear order and constructed in such a way that the associated perturbed gravitational potential is strongly localised to a plane normal to the line of sight at fixed comoving distance from the observer. Now distribute the mass in the plane such that the projected gravitational potential in the plane is purely quadrupolar. For $A>0$, a galaxy in the plane will be stretched such that its major axis is colinear with the line joining the two overdense regions of the quadrupole (the `overdensity axis'). Now consider a circular congruence of CMB photons focusing at the observer, and consider following the trajectory of the congruence backwards towards the surface of last scattering. Lensing by the mass configuration will shear the congruence into an ellipse with major axis aligned with the overdensity-axis. Hence a source on the last scattering surface (e.g. a temperature hot-spot) is stretched along the `underdensity axis', i.e. in the opposite sense to how the tidal field shears a galaxy. The effect is the exact analogy of the GI term in cosmic shear, see Fig.~1 of ~\citet{2004PhRvD..70f3526H}.

\begin{figure}
\centering
\includegraphics[width=\columnwidth]{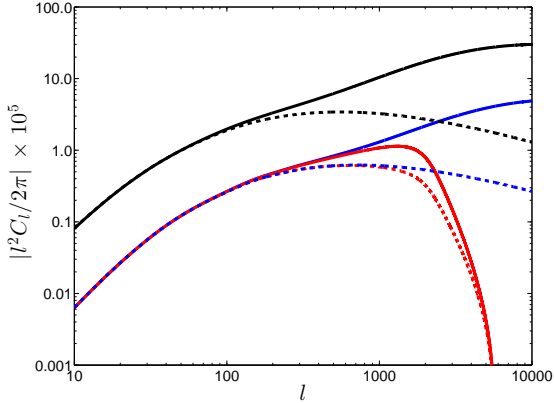}
\caption{The angular cross-spectrum of the CMB lensing E-mode and galaxy lensing E-mode in non-linear (black, upper solid) and linear (black, upper dashed) theory. Also plotted is the absolute value of the angular cross spectrum of the IA E-mode and CMB lensing E-mode in non-linear theory without a cut-off (blue, middle solid) and with a cut-off (red, lower solid), and in linear theory without a cut-off (blue, middle dashed) and with a cut-off (red, lower dashed).}
\label{fig:all_dcl}
\end{figure}

In Fig.~\ref{fig:all_dcl} we plot the absolute value of the IA-CMB lensing cross-spectrum and the `standard' galaxy-CMB lensing cross-spectrum, with and without non-linear corrections. This plot shows that the new term is roughly an order of magnitude lower than the standard term. Despite their different redshift kernels, the cross-spectra exhibit a similar scale dependence, which need not be the case for a more sophisticated IA model.

\begin{figure}
\centering
\includegraphics[width=\columnwidth]{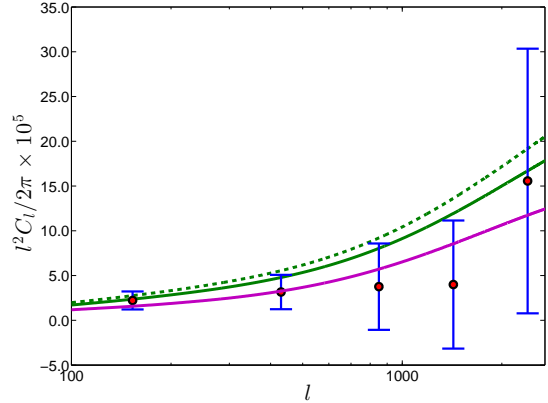}
\caption{The total cross-spectrum of CMB lensing E-modes and observed galaxy ellipticities with (green, upper solid) and without (dashed) an IA term with canonical amplitude, and with an IA model having the best-fitting amplitude of $A=3$ (magenta, lower solid). The data are measurements from~\citet{2013arXiv1311.6200H}, with $1\sigma$ error bars. Errors on neighbouring points are correlated at the percent level.}
\label{fig:cfht_nopoints}
\end{figure}

To compare with the results of~\citet{2013arXiv1311.6200H}, in Fig.~\ref{fig:cfht_nopoints} we plot the non-linear angular cross-power spectrum of CMB lensing and observed galaxy ellipticity, with and without a non-linear IA term without a cut-off. Also plotted are measurements of the cross-power from~\citet{2013arXiv1311.6200H}. Note that neighbouring points on this plot are correlated at the percent level. Including IAs clearly lowers the amplitude of the curve, bringing it closer in line with the data points, and easing the tension with \emph{Planck}. 

To quantify the improved fit to the data, we can model the cross-spectrum as $C_l^{X} = A_p C_l^{\gamma_{CMB} \gamma_{gal}} + C_l^{\gamma_{CMB} \gamma_{IA}}$, where $A_p$ parametrizes the amplitude of the `standard' contribution. The \emph{Planck} best-fit model predicts $A_p=1$, and we will neglect any uncertainties in the source-redshift distribution. In the absence of IAs, \citet{2013arXiv1311.6200H} found that $A_p = 0.61 \pm 0.19$ (with 1$\sigma$ Gaussian errors). Repeating their analysis using the non-linear IA model without a cut-off, we find $A_p = 0.72 \pm 0.19$. Thus, a $2\sigma$ discrepancy becomes approximately a $1.5\sigma$ discrepancy if IAs with the canonical amplitude are included. However, while the inclusion of IAs helps to reconcile theory and observation, both are consistent with the data. Conversely, by fixing $A_p=1$, we find that the data prefer an amplitude for IAs of $A = 3.0 \pm 1.4$, assuming the non-linear model and no cut-off. We plot the best-fitting cross-power spectrum in Fig.~\ref{fig:cfht_nopoints}. Thus, the improvement has only mild statistical significance, but does ease the tension with \emph{Planck}. For example, a model with $A_p=A=1$ has $\chi^2/\nu=0.78$ for $\nu=4$ degrees of freedom, which should be compared to a model with $A_p=1, A=0$, having $\chi^2/\nu = 1.43$.

%To facilitate comparison with the results of~\citet{2013arXiv1311.6200H}, in Fig.~\ref{fig:cfht_nopoints} we plot the non-linear angular cross-power spectrum of CMB lensing and observed galaxy ellipticity, with and without the IA term, plotted in the same way as Figure 4 of~\citet{2013arXiv1311.6200H}. We have also plotted measurements of the cross-power from~\citet{2013arXiv1311.6200H}, read off from their Figure 4. Including IAs clearly lowers the amplitude of the curve, bringing it closer in line with the data points. The IA model is a non-linear model without a cut-off, but the curves are stable to the IA model since the differences are confined to small angular scales.

%Including IAs clearly lowers the amplitude of the curve, bringing it closer in line with the data points in Figure 4 of~\citet{2013arXiv1311.6200H}. The IA model is a non-linear model without a cut-off, but the curves are stable to the IA model since the differences are confined to small angular scales.

%\begin{figure}
%\centering
%\includegraphics[width=\columnwidth]{cfht_nopoints.eps}
%\caption{(Color Online.) The total cross-spectrum of CMB lensing E-modes and observed galaxy ellipticities with (black, solid) and without (blue, dashed) the IA term.}
%\label{fig:cfht_nopoints}
%\end{figure}

%In Fig.~\ref{fig:ratios} we plot the ratio of the IA term and the standard term. The solid lines in this figure are non-linear IA models, and the dashed lines are linear models. Firstly, note that the size of the IA term is roughly 15\% for the $l$-range measured in~\citet{2013arXiv1311.6200H}. Since the measured amplitude of the cross-correlation was found to be roughly 40\% lower than the expected amplitude, and this corresponded to roughly 2$\sigma$, our results suggest that roughly 1$\sigma$ of this discrepancy could be resolved by including IAs with the canonical amplitude. Note that this curve should be scaled linearly for $A\neq 1$.

\begin{figure}
\centering
\includegraphics[width=\columnwidth]{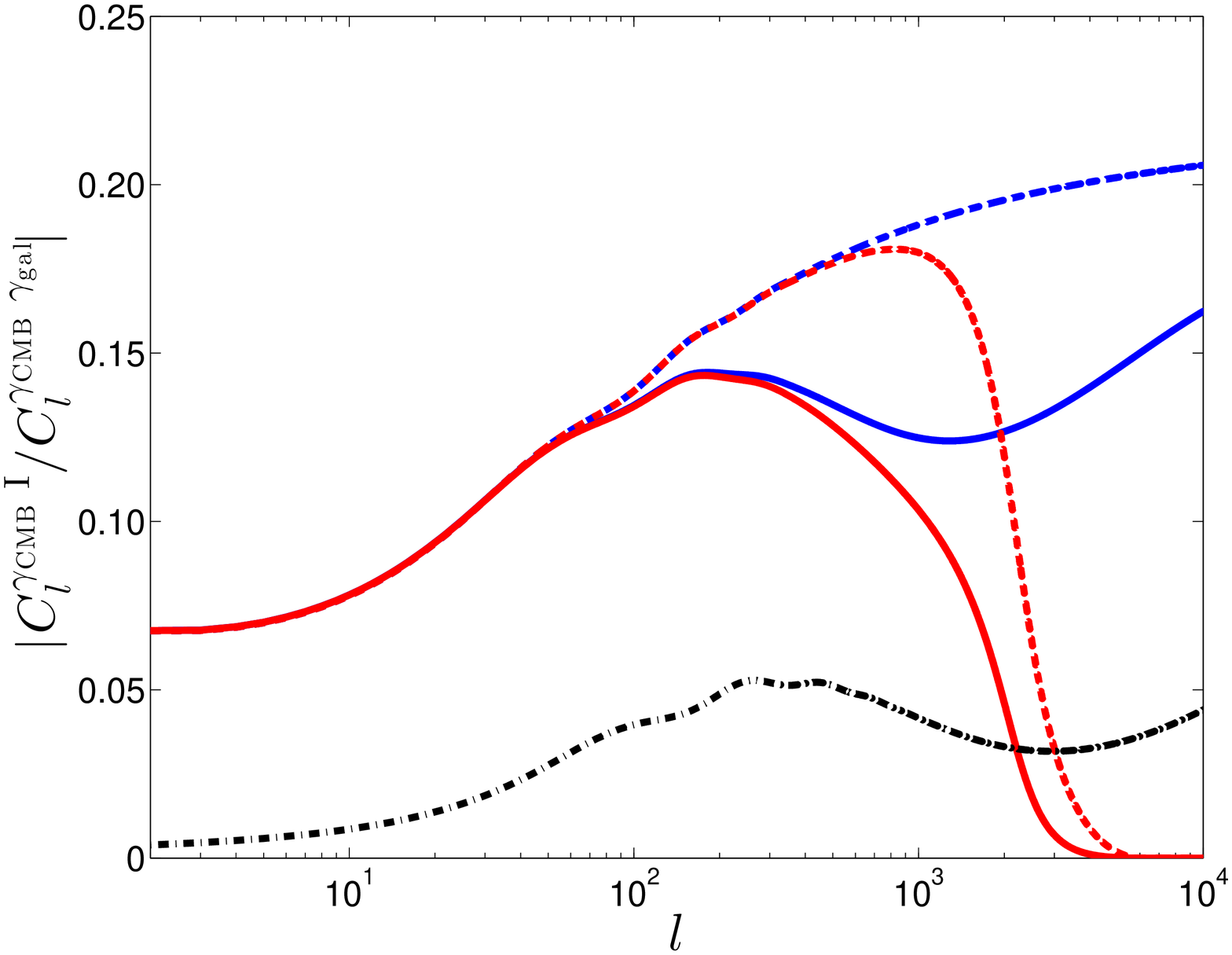}
\caption{The absolute value of the ratio of the cross-power spectrum of CMB lensing and IA E-modes to the cross-power spectrum of CMB lensing and galaxy lensing E-modes. Plotted are the ratios of linear IAs with and without a cut-off to linear standard (red, lower dashed and blue, top dashed respectively) and non-linear IAs with and without a cut-off to non-linear standard (red, lower solid and blue, upper solid respectively). Also plotted is the ratio of a non-linear IA spectrum without a cut-off to a non-linear standard spectrum, for a source-redshift distribution shifted to higher redshift by $\Delta z = 1$ (black, dot-dashed)}
\label{fig:ratios}
\end{figure}

In Fig.~\ref{fig:ratios} we plot the ratio of the IA term and the standard term. The solid lines in this figure are non-linear IA models, and the dashed lines are linear models. Firstly, note that the size of the IA term is roughly 15\% for the $l$-range measured in~\citet{2013arXiv1311.6200H}. Although we have seen that non-linear IA models with the canonical amplitude eases the tension between data and model at the $0.5 \sigma$ level, note that this curve should be scaled linearly for $A\neq 1$. Thus for surveys with a high proportion of galaxies having $A \gtrsim 2$, the new term will be significant. This freedom to fit $A$ arises due to the unknown physics which drives IAs. However, our results indicate that a fully joint analysis of CMB lensing and cosmic shear will be able to place constraints on $A$ that improve upon what may be achieved with cosmic shear alone.

Including non-linear effects in the IA and lensing models reduces the significance of the IA term. This is because the galaxy lensing shear picks up contributions from matter at lower redshifts than the IA shear, which is confined to the higher galaxy redshifts. For a fixed $l$, lower redshifts correspond to larger wavenumbers, which receive more enhancement from non-linearities. Thus the denominator in the ratio plotted in Fig.~\ref{fig:ratios} is boosted by more than the numerator, and so the ratio is suppressed.

Finally, to investigate the dependence on the source-redshift distribution, in  Fig.~\ref{fig:ratios} we also plot the ratio of the IA term to the standard term in non-linear theory without a cut-off for the $n_z$ distribution of equation~(\ref{eq:zdist}) shifted to higher redshift by a constant translation of $\Delta z = 1$. Note that the small-scale variation in the distance distribution induced by this operation decreases the angular scale where the Limber approximation is expected to be accurate. The relative importance of the IA term now decreases. This can be understood by noting that the galaxy lensing E-mode increases in amplitude due to the enhanced lensing kernel for higher-redshift sources. In contrast, the equivalent distance kernel for the IA term is the source distribution $n_r$, which does not receive a significant enhancement.
%
%\begin{figure}
%\centering
%\includegraphics[width=\columnwidth]{rat_hiz.eps}
%\caption{Same as in Fig.~\ref{fig:ratios}, for the non-linear and no cut-off IA model, for the distribution in Eq.~\eqref{eq:zdist} (solid) and the same distribution shifted to higher redshift by $\Delta z = 1$ (dashed).}
%\label{fig:rat_hiz}
%\end{figure}
%
\section{Conclusions}
\label{sec:concs}

In this work, we have shown how intrinsic galaxy alignments can contribute significantly to the cross-correlation of CMB weak lensing and galaxy weak lensing. For a simple tidal field model with a canonical amplitude, the IA term can reduce the cross-power by roughly 15\% for the CS82 redshift distribution. Reanalysing measurements of the cross-power from~\citet{2013arXiv1311.6200H} shows that including IAs with the canonical amplitude recovers roughly $0.5 \sigma$ of the $2 \sigma$ discrepancy between the data and the best-fit \emph{Planck} cosmology. Indeed, all the discrepancy may be removed by fixing the IA amplitude to $A=3$, although the improvement is mild. As a caveat to this however, the CFHTLenS analysis of~\citet{2013MNRAS.432.2433H} found a value for $A$ consistent with zero. This can be partly understood by the fact that the CFHTLenS source-galaxy population is dominated by late-type galaxies, which have low values of $A$ in the linear alignment model. However, a more satisfactory explanation of IAs for such galaxies is provided by tidal-torque theory, which predicts no `GI' term at linear order~\citep{2004PhRvD..70f3526H}, and hence no contribution to the cross-correlation with CMB lensing at linear order. If the galaxy sample used in~\citet{2013arXiv1311.6200H} is similarly dominated by late-type galaxies, we would not expect the new term to be significant. Nonetheless, our analysis has shown that IAs can be very important for early-type galaxies, for which the simple linear model is expected to apply.

%In this work, we have shown how intrinsic galaxy alignments can contribute significantly to the cross-correlation of CMB weak lensing and galaxy weak lensing. For a simple tidal field model with a canonical amplitude, the IA term can reduce the cross-power by roughly 15\% for the CS82 redshift distribution. I thus has the potential to reconcile the recent measurements of the cross-power in~\citet{2013arXiv1311.6200H} with the best-fitting model found from \emph{Planck}. As a caveat to this however, the CFHTLenS analysis of~\citet{2013MNRAS.432.2433H} found a value for $A$ consistent with zero. This can be partly understood by the fact that the CFHTLenS source galaxy population is dominated by late-type galaxies, which have low values of $A$ in the linear alignment model. However, a more satisfactory explanation of IAs for such galaxies is provided by tidal-torque theory, which predicts no `GI' term at linear order~\citep{2004PhRvD..70f3526H}, and hence no contribution to the cross-correlation with CMB lensing at linear order. If the galaxy sample used in~\citet{2013arXiv1311.6200H} is similarly dominated by late-type galaxies, we would not expect the new term to be significant. Nonetheless, our analysis has shown that IAs can be very important for early-type galaxies, for which the simple linear model is expected to apply.

% Furthermore, the amplitude $A$ of IAs could be much greater than unity with this redshift distribution, as evidenced by the 68\% upper limit of $A=6$ found in~\citet{2008A&A...479....9F}.

Even if the amplitude of IAs is not sufficient to fully reconcile the measurement, we have demonstrated that it should at least be included in models of the cross-correlation, and could even be constrained by such measurements. This is an attractive possibility, as the precisely known redshift of the last scattering surface, as well as the cleanness of CMB weak lensing as a probe, could allow a fully joint analysis to `self-calibrate' the amplitude of IAs in cosmic shear surveys. This would almost certainly aid attempts to mitigate the many systematics in cosmic shear with the use of CMB lensing, as recently suggested in~\citet{2012ApJ...759...32V, 2013ApJ...778..108V,2013arXiv1311.2338D}, although the uncertainty in the source-redshift distribution, the time-dependence of the IA amplitude, and uncertainties in the small-scale IA mechanism would need to be accounted for. We intend to investigate this intriguing possibility in a forthcoming work.

Further potential uses of this signal include improved delensing for CMB B-mode detection~\citep{2007PhRvD..76l3009M} and the use of IAs to probe tidal fields~\citep{2013JCAP...12..029C}. In any case, our analysis has demonstrated that IAs should be included in any future study of this particular cross-correlation.

\section{Acknowledgments}
AH thanks Anthony Challinor and Catherine Heymans for helpful discussions, and Nick Hand for providing the CS82-ACT data. AH is supported by an STFC Consolidated Grant. A few days after submitting this letter, Troxel \& Ishak (arXiv:1401.7051) appeared, which deals with similar issues to those considered here. Their work is complementary to ours, and the results agree where there is overlap.

\bibliographystyle{mn2e_fix}
\bibliography{references}

\end{document}